\documentclass[aps,prb,preprint,11pt,showkeys,nobibnotes,floatfix]{revtex4}

\usepackage{amsmath,graphicx}

\begin{document}



\title{Cyclotron Resonance of Itinerant Holes
in Ferromagnetic InMnAs/GaSb Heterostructures}

\author{G. A. Khodaparast}
\author{J. Kono}
\email[Please send all correspondence to: Prof. Junichiro Kono,
Rice University, ECE Dept., MS-366, P.O. Box 1892,
Houston, TX 77251-1892, U.S.A.  Phone: +1-713-348-2209.
E-mail:]{kono@rice.edu}
\affiliation{Department of Electrical and Computer Engineering,
Rice Quantum Institute, and Center for Nanoscale Science
and Technology, Rice University, Houston, Texas 77005}

\author{Y. H. Matsuda}
\thanks{Present address: Department of Physics, Okayama University,
Okayama, Japan.}
\author{T. Ikaida}
\author{S. Ikeda}
\author{N. Miura}
\affiliation{Institute for Solid State Physics, University of
Tokyo, Kashiwa, Chiba 277-8581, Japan}

\author{T. Slupinski}
\affiliation{Institute of Experimental Physics, Warsaw University,
Hoza 69, 00-681 Warsaw, Poland}
\author{A. Oiwa}
\author{H. Munekata}
\affiliation{Imaging Science and Engineering Laboratory,
Tokyo Institute of Technology, Yokohama, Kanagawa 226-8503, Japan}

\date{\today}

\begin{abstract}
We report the first observation of hole cyclotron resonance (CR) in
ferromagnetic InMnAs/GaSb heterostructures both in the high-temperature
paramagnetic phase and the low-temperature ferromagnetic phase.  We
clearly resolve two resonances that exhibit strong temperature
dependence in position, linewidth, and intensity.
We attribute the two resonances to the so-called fundamental CR transitions
expected for delocalized holes in the valence band in the magnetic
quantum limit, demonstrating the existence of $p$-like itinerant holes
that are describable within the Luttinger-Kohn effective mass theory.



\end{abstract}

\maketitle
\section{INTRODUCTION}
The interaction of free carriers with localized spins plays an
important role in a variety of magnetic and many-body phenomena in metals.
\cite{anderson,kittel,hewson}  Carriers in the vicinity of a magnetic ion
are magnetized, which in turn leads to an indirect exchange interaction
between magnetic ions.  The discovery of carrier-induced ferromagnetism
in magnetic III-V semiconductors\cite{ohno,munekata,ohno1} has not
only opened up new device opportunities but also provided a novel
material system in which to study the physics of itinerant carriers
interacting with localized spins.  Various theoretical models have
been proposed but the microscopic mechanism is still a matter of controversy.
\cite{dietl,konig,litvinov,sarma}
One of the open questions is the nature of the carriers mediating
the exchange interaction between Mn ions, i.e., whether they
reside in the impurity band ($d$-like), the delocalized
valence bands ($p$-like), or some type of mixed states.

In this paper we describe our observation of hole cyclotron resonance (CR)
in ferromagnetic InMnAs/GaSb heterostructures, unambiguously demonstrating
the existence of delocalized $p$-like carriers.  In addition, to our
knowledge, this is the first study of CR in any ferromagnetic system
covering temperature ranges both below and above the Curie temperature
($T_c$).\cite{grimes,sievers}  CR is a direct and accurate method for
determining the effective masses of carriers (i.e., the curvature of the
energy dispersion) and therefore the nature of the carrier states.  In all
the samples studied, we observed two pronounced resonances.  Both lines
exhibited unusual temperature dependence in their position,
intensity, and width.  The lower-field resonance showed an abrupt
reduction in linewidth with a concomitant decrease in resonance magnetic
field slightly above $T_c$.  The higher-field line, which was absent
at room temperature, suddenly appeared above $T_c$, rapidly grew in
intensity with decreasing temperature, and became comparable to the
lower-field resonance at low temperatures.  We ascribe these lines to
the two fundamental CR transitions expected for delocalized holes in
the valence band of a Zinc-Blende semiconductor in the magnetic quantum limit.
We take this as evidence for the existence of a large density of delocalized
$p$-like holes in these ferromagnetic systems.

\section{SAMPLES STUDIED}
Three samples were studied.  They were InMnAs/GaSb single
heterostructures, consisting of 9-30 nm of InMnAs with Mn content $\sim$
9\% and a 600-800 nm thick GaSb buffer grown on semi-insulating
GaAs (100) substrates.  They contained high densities
($\sim$ 10$^{19}$ cm$^{-3}$) of holes provided by the Mn acceptors.
The samples were grown by low temperature molecular beam epitaxy.
The growth conditions have been described previously.\cite{tom}
Unlike the paramagnetic $n$-type\cite{zudov} and $p$-type\cite{matsuda1}
InMnAs films we studied earlier, the samples in the present work showed
ferromagnetism with $T_c$ ranging from 30 K to 55 K.
The magnetization easy axis was perpendicular to the epilayer due to
the strain-induced structural anisotropy caused by the lattice mismatch
between InMnAs and GaSb (InMnAs was under tensile strain).  The energy
level structure of InMnAs/GaSb is complicated and known to have a
'broken gap' type-II configuration.  The hole densities and mobilities
($\sim$ 10$^{19}$ cm$^{-3}$ and $\sim$ 300 cm$^{2}$/vs, respectively)
were estimated from room temperature Hall measurements and the Curie
temperatures were determined by magnetization measurements.  Sample 1
was annealed at 250 $^{\circ}$C after growth, which increased the $T_c$
by $\sim$ 10 K.\cite{hayashi,potashnik}

\section{EXPERIMENTAL TECHNIQUES}
We performed infrared (IR) CR measurements using ultrahigh pulsed
magnetic fields ($\le$ 150 Tesla) generated by the single-turn coil
technique.\cite{nakao}  The magnetic field, applied along the growth
direction, was measured by a pick-up coil around the sample, which was
placed inside a continuous flow helium cryostat.  We used IR laser
beams with wavelengths of 10.6 $\mu$m, 10.2 $\mu$m, 9.25 $\mu$m
(CO$_2$ laser), and 5.527 $\mu$m (CO laser).  We circular polarized
the IR radiation using a CdS quarter-wave plate.  The transmitted
radiation through the sample was collected using a fast
liquid-nitrogen-cooled HgCdTe photovoltaic detector.
A multi-channel digitizer placed in a shielded room recorded the signals
from the pick-up coil and the detector.  Although the coil breaks in
each shot, the sample and pick-up coil remain intact, making it possible to
carry out detailed temperature and wavelength dependence studies
on the same specimen.  Since the transmission signal was recorded
during both the up and down sweeps, each resonance was observed twice
in a single pulse.  This allowed us to check the reproducibility of observed
absorption peaks and to make sure that the spectra were free from any
slow heating effects.


\section{EXPERIMENTAL RESULTS AND DISCUSSION}
Figures 1(a) and 1(b) show the transmission of the 10.6 $\mu$m beam
($\hbar \omega=0.117\ \mbox{eV}$) through sample 1 ($T_c$ = 55 K) and
sample 2 ($T_c$ = 30 K), respectively, at various temperatures as a
function of magnetic field.  The beam was hole-active circularly polarized.
In Fig. 1(a), from room temperature down to about 80 K, a broad resonance
feature (labeled 'A') is observed with almost no change in intensity,
position, and width with decreasing temperature.  However, with further
decreasing temperature, we observe quite abrupt and dramatic changes in
the spectra.  First, a significant reduction in linewidth and a sudden
shift to a lower magnetic field occur simultaneously.  Also, it rapidly
increases in intensity with decreasing temperature.  In addition, a
second feature (labeled 'B') suddenly appears around 125 Tesla, which
also grows rapidly in intensity with decreasing temperature and saturates,
similar to feature A.  At low temperatures, both features A and B do not
show any shift in position.  Essentially the same behavior is seen for
sample 2 in Fig. 1(b).

From Lorentzian fits to the CR data, we deduced the values for the
cyclotron mass, density, and mobility.
The hole cyclotron masses obtained for this wavelength (10.6 $\mu$m)
for peaks A and B are 0.051$m_0$ and 0.12$m_0$, respectively, where
$m_0$ is the mass of free electrons in vacuum.  The
obtained densities and mobilities for feature A are plotted in Figs. 2(a)
and 2(c) for samples 1 and 2, respectively, together with the temperature
dependence of the magnetization $M$ in (b) and (d).
It is interesting to note that the estimated CR mobility at the lowest
temperature of our experiments ($\sim$ 15 K) is surprisingly high, i.e.,
4$-$5 $\times$ 10$^3$ cm$^2$/Vs.  This kind of high CR mobility is totally
incompatible with the low DC mobilities deduced from Hall measurements,
which are $\sim$ 300 cm$^2$/Vs and sometimes even decrease with
decreasing temperature.  This suggests that in these
ferromagnetic semiconductors a DC mobility is not a good
quantity for assessing the CR observability condition, i.e.,
$\omega_c\tau = B\mu > 1$, where $\omega_c = eB/m^*$ is the cyclotron
frequency, $\tau$ is the scattering time, and $\mu$ is the mobility.

Figure 3(a) shows low temperature
CR traces for the three samples taken with 10.6 $\mu$m radiation.
Both features A and B are clearly
observed but their intensities and linewidths vary from sample
to sample, depending on the density, mobility, and thickness.
The observed unusual temperature dependence is not specific to this
particular wavelength.
Figure 3(b) displays the wavelength dependence of the CR
spectra for sample 2.  We can see that both lines shift
to higher magnetic fields with decreasing wavelength (i.e.,
increasing photon energy), as expected.
Figures 3(c) and 3(d) show data at different temperatures
for sample 1 measured with 9.25 $\mu$m and 5.52 $\mu$m
radiation, respectively.  The temperature dependence observed
at these shorter wavelengths is
similar to what was observed at 10.6 $\mu$m.  All these data
confirm the universality of the effects we observed.



The clear observation of CR indicates that there are delocalized holes
in these ferromagnetic samples.  This is in agreement with our CR
measurements on low-$T_c$ $p$-type InMnAs films in the paramagnetic
phase,\cite{matsuda1} which showed similar two resonance CR spectra
although the resonances were much broader, temperature dependence
was weaker, and the resonance positions were slightly (< 10\%) lower than
the present heterostructure samples.  We attribute the resonances
to the two CR transitions expected in the magnetic quantum
limit (the so-called 'fundamental' transitions\cite{bruce}), one being
heavy-hole-like ($m_J = -3/2$) and the other light-hole-like ($m_J = -1/2$).
The initial states of these transitions are the two lowest ($n=-1$)
Landau levels in the valence band, and the corresponding cyclotron
masses at $k_z = 0$ are given by $(\gamma_1 \pm \bar{\gamma})^{-1}m_0$
within the spherical approximation based on a 4 $\times$ 4 Luttinger
Hamiltonian,\cite{bruce} where $k_z$ is the wavenumber in the magnetic
field direction and $\gamma_1$ and $\bar{\gamma}=(\gamma_2 + \gamma_3)/2$
are Luttinger parameters.
More detailed calculations based on an eight-band effective mass model
including finite $k_z$ effects successfully reproduced these two
resonances for pure InAs and paramagnetic InMnAs films.\cite{gary}
We believe that the $\sim$10\% mass enhancement in the
heterostructure samples compared to the bulk films is due to
quantum confinement (layer thickness only $\sim$ 9 nm) plus non-parabolicity.

We anticipate that the experimental findings presented here will
stimulate interest in the problem of the cyclotron resonance of
itinerant carriers in ferromagnets, and more theoretical studies
will be carried out to explain, in particular, the unusual
temperature dependence we observed.  We currently have no adequate
explanation for the abrupt changes in mass, width and intensity of
CR.  The rapid line narrowing of the lower-field line as well as the
sudden appearance of the higher-field line is equally striking.
The ferromagnetic order should split the valence bands {\em even in the
absence of a magnetic field}, which should also strongly modify
Landau and Zeeman quantization at high fields.
It is important to emphasize that the temperature at which the
significant spectral changes start to appear ($T_c^*$) is consistently
higher than the Curie temperature ($T_c$) in all three samples.
This fact could be explainable in light of a recent Monte Carlo study
by Schliemann {\em et al.},\cite{schliemann} which showed that
short-range magnetic order and finite {\em local} carrier spin
polarization are present for temperatures substantially higher
than $T_c$.  Any such order should result in modifications in band structure,
which in turn modify CR spectra.

\section{SUMMARY}
We observed the cyclotron resonance of itinerant holes in ferromagnetic
InMnAs/GaSb heterostructures both above and below $T_c$.
We observed two pronounced resonances that were strongly temperature
dependent in position, width and intensity.  We attribute these transitions
to the 'fundamental' light hole and heavy hole cyclotron resonance,
the observation of which clearly demonstrates that there are delocalized
$p$-like holes in InMnAs.
This important information on the carrier states should provide
new insight into the microscopic mechanism of carrier-induced
ferromagnetism in this family of magnetic semiconductors.

\section{ACKNOWLEDGEMENTS}

We gratefully acknowledge support from DARPA MDA972-00-1-0034,
NSF DMR-0049024, DMR-0134058 (CAREER), and NEDO.
 

\newpage


\begin{figure}
\includegraphics[scale=1.2]{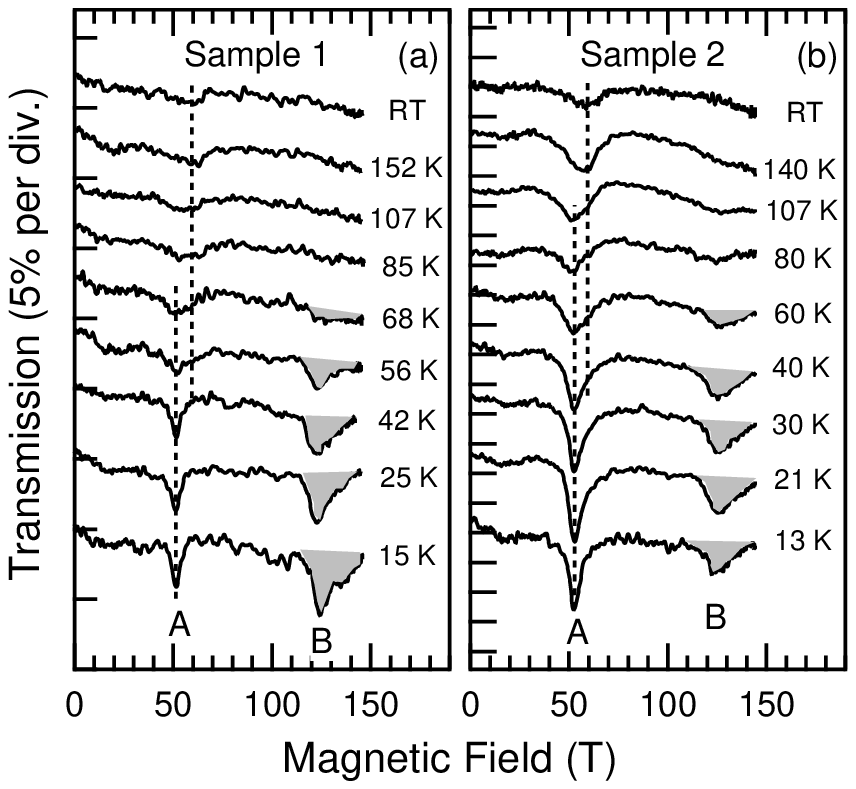}
\caption{Cyclotron resonance spectra for (a) sample 1 and (b) sample 2.
The transmission of hole-active circularly polarized 10.6 $\mu$m radiation
($\hbar \omega=0.117\ \mbox{eV}$) is plotted as a function of magnetic
field at different temperatures.  Both samples show two strongly
temperature-dependent features, labeled A and B, whose origins are
discussed in the text.
}
\label{fig1}
\end{figure}
\begin{figure}[tbp]
\includegraphics[scale=1.2]{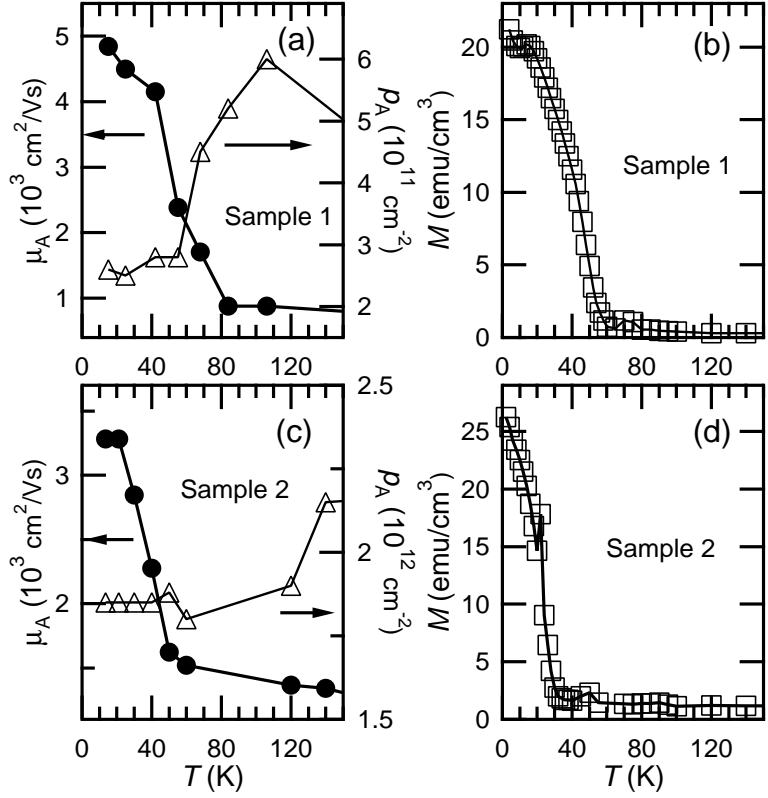}
\caption{(a) and (c): Hole mobilities ($\mu_{A}$) and densities
($p_{A}$) vs. temperature ($T$), deduced from the integrated
intensity and linewidth of feature A in Fig. 1.
(b) and (d): Magnetization ($M$) vs. temperature for samples 1 and 2,
obtained by SQUID measurements with a magnetic field of 0.5 mT
applied along the growth direction.
}
\label{fig2}
\end{figure}
\begin{figure}[tbp]
\includegraphics[scale=1.2]{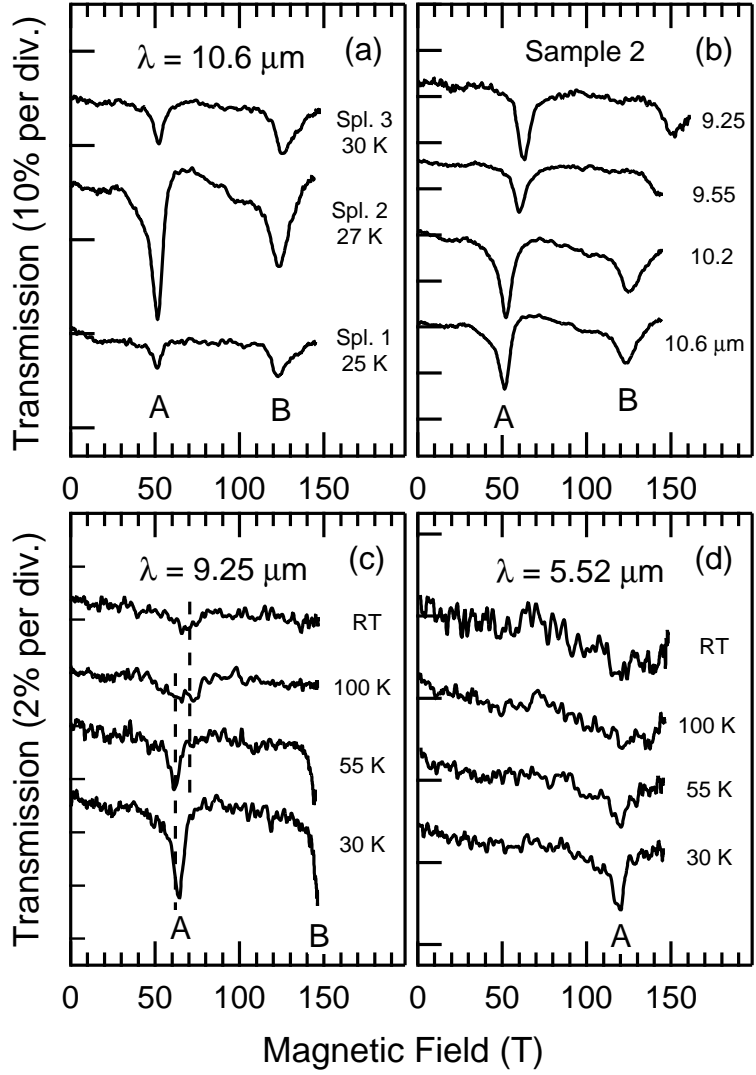}
\caption{(a) Low temperature CR spectra for the three samples at
10.6 $\mu$m.  (b) Wavelength dependence of the CR spectra for
sample 2 at 27 K.  (c) CR spectra for sample 1
at different temperatures at 9.25 $\mu$m.  (d) CR spectra
for sample 1 at 5.52 $\mu$m.}
\label{fig3}
\end{figure}

\end{document}